\documentclass[fleqn,12pt,twoside]{article}
\usepackage{espcrc1}

\newcommand{\AmS}{{\protect\the\textfont2
  A\kern-.1667em\lower.5ex\hbox{M}\kern-.125emS}}

\title{New Developments in Treacherous Points of Light-Front Dynamics}

\author{Chueng-Ryong Ji\address[MCSD]{Department of Physics,
North Carolina State University, Raleigh, NC 27695-8202, USA}%
\thanks{Supported by Department of Energy under the contract 
DE-FG02-96ER40947.},
Bernard L. G. Bakker\address{Department of Physics and Astronomy,
Vrije Universiteit, Amsterdam, The Netherlands}
and
Ho-Meoyng Choi\address{Department of Physics, Teachers College,
Kyungpook National University, Daegu, Korea 702-701}%
\thanks{
Supported by Korean Research Foundation under the contract
KRF-2005-070-C00039.}}
       
\begin{document}

\maketitle

\begin{abstract}

Light-front dynamics(LFD) plays an important role in hadron
phenomenology.  The last few years, however, it has been emphasized
that treacherous points such as zero-mode contributions should be taken
into account for successful LFD applications to hadron phenomenology.
We discuss some examples of treacherous points and present new progress
made the last few years to handle them correctly.

\end{abstract}

\section{INTRODUCTION}

Light-front dynamics (LFD) provides a unified framework to analyze
various experimental measurements such as generalized parton
distributions (GPDs) and single spin asymmetries (SSA) at JLab and DESY
(Hermes)~\cite{GPV}, B-decays at SLAC (BaBar) and KEK
(Belle)~\cite{JC-BPhysics} as well as quark gluon plasma (QGP)
production at BNL (RHIC) and CERN (ALICE)~\cite{Duke}, etc..  Owing to
the rational energy-momentum dispersion relation, LFD has distinct
features compared to other forms of Hamiltonian dynamics. In
particular, the vacuum fluctuations are suppressed and the kinematic
generators are proliferated in LFD. Overall, these distinct features
can be regarded as advantageous in hadron phenomenology.  However, LFD
implies also treacherous points as one may realize from the
significance of zero-mode contributions even in the good (+) current
analyses~\cite{BCJ-spin1}. It has also been shown that the common belief
of equivalence between manifestly covariant calculations and naive
light-front (LF) calculations is not always realized~\cite{BJ-endpoint}
unless treacherous points are well taken care of~\cite{BDJM}. Thus,
careful investigations of treacherous points and judicious ways of
handling them should be imperative for LFD to be distinctively
useful compared to other forms of Hamiltonian dynamics~\cite{Dirac}.

In this presentation, we discuss an example that requires the inclusion
of the arc contribution in the light-front energy contour integration
in order to achieve equivalence between the LFD result and the
manifestly covariant result. For a concrete example of a zero-mode
contribution, we also summarize the Standard Model analysis of the
vector anomaly in the CP-even form factors of $W^{\pm}$ gauge bosons.
We then present a power counting method to correctly pin down which
hadron form factors receive a zero-mode contribution and which ones
do not.

\section{ARC CONTRIBUTION}
It is well known that spurious divergences appear in LF calculations
that do not appear when the same amplitudes are computed in the
standard equal-time formulation.  This has been taken to mean that the
equivalence between the LF and manifestly covariant formalisms is not
complete.  These divergences have been regulated with various methods
including the principal-value prescription, the Mandelstam-Leibbrandt
prescription, cutoffs, smearing, and BPHZ-like differentiation and
reintegration.  We encountered an example of such a divergence in one
of our previous works~\cite{BJ-endpoint}.  There, an end-point
singularity was seen in the (1+1)-dimensional calculation of the
pseudoscalar (and scalar) elastic form factor.  Recently, we have
shown~\cite{BDJM} that the integrand in this case does not vanish
sufficiently fast as the LF energy ($k^-$) goes to infinity and the arc
contribution is nonzero so that it must be included in order to obtain
the correct result.  We also showed~\cite{BDJM} that point contributions
occur when the integration contour in the $k^-$ plane crosses a moving
singularity.

The key point of the arc contribution may be seen from the following 
matrix element:
\begin{equation}
<p'|J^-|p> = 2N \int \frac{dk^+ dk^-}{(2\pi)^2}
\frac{(m^2- k^+k^-)(k^- - p^- -p'^-) - k^+ p^- p'^-}
{k^+(k^+ - p^+)(k^+ - p'^+)(k^- - k_1^-)(k^- - k_2^-)(k^- - k_3^-)},
 \label{eq.001}
\end{equation}
where $k_1^- = \frac{m^2-i\epsilon}{k^+}, k_2^- = p^-
+\frac{m^2-i\epsilon}{k^+-p^+},~k_2^- = p^-
+\frac{m^2-i\epsilon}{k^+-p^+}$.  We see from this equation that the
numerator has a term proportional to $(k^-)^2$ while the highest power
of $k^-$ in the denominator is 3. The numerator has also $dk^-$, so, if
we set $k^- = R \exp{i\theta}$ to compute the contribution from the arc
of the $k^-$ contour integration, then we find that the term
proportional to $(k^-)^2$ in the numerator has no $1/R$ suppression
factor and can contribute to the form factor calculation.

For the $k^+ < 0$ region, all three poles $k_1^-,k_2^-,k_3^-$ are in the 
upper half-plane and thus by closing the contour in the lower half-plane 
we get
\begin{equation}
\int_{-\infty}^{+\infty} dk^- = - \int_{\rm arc} dk^- .
\end{equation}
From this, we find the following contribution for the $k^+ < 0$ region:
\begin{equation}
<p'|J^-|p> = \frac{2N}{(2\pi)^2} \int_{-\infty}^0 dk^+ \frac{-i\pi}{(k^+ - p^+)(k^+ -
p'^+)}.
\end{equation}
 
Similary, we find the arc-contributions from the other regions;
$0<k^+<p^+$, $p^+<k^+<p'^+$, and $k^+>p'^+$. Summing all arc-contributions
for the entire $k^+$ region, we get
\begin{eqnarray}
<p'|J^-|p> &=& \frac{i2N\pi}{(2\pi)^2}\left[-\int_{-\infty}^0 -\int_0^{p^+}
+\int_{p^+}^{p'^+} +\int_{p'^+}^{\infty} \right]
 \frac{dk^+}{(k^+-p^+)(k^+-p'^+)}\\
\nonumber
&=& \frac{iN}{2\pi}\left[-\int_{-\infty}^{p^+} + \int_{p^+}^{p'^+}
+\int_{p'^+}^{\infty} \right] \frac{dk^+}{(k^+-p^+)(k^+-p'^+)}.
\end{eqnarray}
Here, we can show that the contributions from $\int_{-\infty}^{p^+}$ and
$\int_{p'^+}^{\infty}$ cancel each other exactly.
Thus, we find the following arc-contribution:
\begin{equation}
<p'|J^-|p> = \frac{iN}{2\pi}\int_{p^+}^{p'^+} \frac{dk^+}{(k^+-p^+)(k^+-p'^+)}.
\end{equation}  
This contribution leads to the correct result for the physical form
factor removing the discrepancy between the $J^+$ and $J^-$ calculations.

\section{VECTOR ANOMALY}

Anomalies betray the true quantal character of a quantized field
theory.  A brief historical remark on the anomaly associated with the
fermion-triangle loop was made in Ref.~\cite{Ji-Bakker} and our
calculation of the vector anomaly has been presented in detail for the
Standard Model (SM) in Refs.~\cite{BJnew,JBC}. In conjunction with the LF
singularities~\cite{BBJ} presented by Ben Bakker in this conference, we
briefly summarize the vector anomaly results including the zero-mode
contribution in LFD.

The Lorentz-covariant and gauge-invariant CP-even electromagnetic
$\gamma W^+ W^-$ vertex is defined~\cite{BGL,CN} by
\begin{eqnarray}
\Gamma^\mu_{\alpha\beta} = i\,e \left\{ A[ (p+p')^\mu g_{\alpha\beta}
 +2(g^\mu_\alpha q_\beta - g^\mu_\beta q_\alpha)] 
 + (\Delta \kappa)( g^\mu_\alpha q_\beta - g^\mu_\beta q_\alpha)
 + \frac{\Delta Q}{2M^2_W} (p+p')^\mu q_\alpha q_\beta\right\}, 
\label{eq.II.010}
\end{eqnarray}
where $p(p')$ is the initial(final) four-momentum of the $W$ gauge boson
and $q=p'-p$.  Here, $\Delta \kappa$ and $\Delta Q$ are the anomalous
magnetic and quadrupole moments, respectively. Beyond the tree level, 
\begin{eqnarray}
A = F_1(Q^2), \quad -\Delta \kappa = F_2(Q^2) + 2 F_1(Q^2), 
\quad -\Delta Q = F_3(Q^2),
\end{eqnarray}
where $F_1, F_2$ and $F_3$ are the usual electromagnetic form factors 
for the spin-1 particles~\cite{BJ02}.

While the anomalous quadrupole moment $\Delta Q$ (or $F_3(Q^2)$) is
found to be completely independent of the regularization methods as it
must be, we find that the anomalous magnetic moment $\Delta \kappa$ (or
$F_2(Q^2) + 2F_1(Q^2)$) differs by some fermion-mass-independent
constants depending on the regularization methods~\cite{BJnew,JBC}.
Unless the fermion-mass-independent differences are completely
cancelled, a unique prediction of $\Delta \kappa$ would be impossible.
Within the SM, they completely cancel out owing to the anomaly free
condition, i.e. the zero-sum of the charge factors ($\sum_f Q_f =0$) in each
generation.

In LFD, we compute the form factors using the following helicity
amplitudes $G^+_{h'h} = \epsilon^{*}_{h'} \cdot \Gamma^{+} \cdot
\epsilon_{h}$ for the initial and final polarization vectors
$\epsilon_h$ and $\epsilon^{*}_{h'}$, respectively, in the $q^+ = q^0 +
q^3 =0$ frame,
\begin{eqnarray}
G^+_{++}&=& 2p^+(F_1 + \eta F_3), 
G^+_{+0} = p^+\sqrt{2\eta} (2F_1 + F_2 + 2\eta F_3),\nonumber\\
G^+_{+-}&=&-2p^+\eta F_3,
G^+_{00} = 2p^+(F_1 - 2\eta F_2 - 2\eta^2 F_3),
\label{eq.II.100}
\end{eqnarray}
where $\eta = Q^2/4M^2_W$.  In the $q^+=0$ frame, one might expect that
the non-valence contribution is absent since the integration range for
the non-valence amplitude is shrunk to zero. However, this is not the
case as we pointed out in Ref.~\cite{BCJ-spin1}. Calling the non-zero
contribution from the non-valence part in the $q^+ = 0$ frame the
zero-mode, we find that only the helicity zero-to-zero amplitude
$G^+_{00}$ receives a zero-mode contribution given by
\begin{eqnarray}
\left( G^+_{00} \right)_{\rm z.m.}
= \frac{g^2 Q_f p^+}{2\pi^3 M^2_W} \int^1_0 dx \int d^2 {\vec k}_\perp
\frac{{\vec k}^2_\perp + m^2_1 -x(1-x)Q^2}{{\vec k}^2_\perp + m^2_1 +x(1-x)Q^2}
\neq 0.
\end{eqnarray}
The zero-mode contribution to $G^+_{00}$ is essential, because the
unwelcome divergences from the valence part due to the terms with a
power of the transverse momentum such as $(k_\perp^2)^2$ are precisely
cancelled by the same terms with the opposite sign from the zero-mode
contribution. One may attempt to remove the zero-mode using
Pauli-Villars regularization.  However, we find that such an artificial
removal of the zero-mode makes the LF calulcation impossible
introducing uncontrollable divergences. The details of our calculations
were presented in Ref.~\cite{BJnew}.

The symptom of vector anomaly in LFD also appears as the violation of
the rotation symmetry or the angular momentum conservation ({\it i.e.}
angular condition~\cite{CJ}).  This appearance is drastically different
from the case of the manifestly covariant calculation. However, the
anomaly-free condition ($\sum_f Q_f = 0$) in the SM again removes the
difference and restores the rotation symmetry and the angular momentum
conservation in LFD.

\section{POWER COUNTING METHOD}

Jaus~\cite{Jaus} proposed a covariant LF approach involving the
lightlike four vector $\omega^\mu (\omega^2 = 0)$ as a variable and
developed a way of finding the zero-mode contribution to remove the
spurious amplitudes proportional to $\omega^\mu$.  Our formulation,
however, is intrinsically distinguished from this $\omega$-dependent
formulation since it involves neither $\omega^\mu$ nor any unphysical
form factors. Our method of finding the zero-mode contribution is a
direct power-counting of the longitudinal momentum fraction in the
$q^+\to 0$ limit for the off-diagonal elements in the Fock-state
expansion of the current
matrix~\cite{BCJ-spin1,BCJ-spin01,CJ-spin01,CJnew}. This power-counting
method is straightforward, since the longitudinal momentum fraction is
one of the integration variables in the LF matrix elements ({\it i.e.}
helicity amplitudes) and the behaviors of the longitudinal momentum
fraction in the integrand are known from the helicity amplitudes.

For spin-1 electroweak form factors and using the rather simple (manifestly
covariant) vertex $\Gamma^\mu =\gamma^\mu$, both Jaus and we agree on
the absence of zero-mode contributions.  However, Jaus and we do not
agree when $\Gamma^\mu$ is extended to the more phenomenologically
accessible ones given by
\begin{equation}
\label{eq1}
\Gamma^\mu=\gamma^\mu -\frac{(k+k')^\mu}{D},
\end{equation}
where $k$ and $k'$ are the relative four momenta for the two
constituent quarks.  Although Jaus's calculation and our calculation
used the same denominator $D$ in Eq.~(\ref{eq1}), they led to different
conclusions in the analysis of the zero-mode contribution.  Even if $D$
is chosen in such a way as to get the manifestly covariant
$\Gamma^\mu$, the difference in the conclusions doesn't go away.

For an example of the weak transition form factors between the
pseudoscalar and vector mesons, Jaus~\cite{Jaus} concluded that
the form factor $A_1(q^2)$[or $f(q^2)$] receives a zero-mode
contribution.  We do not agree with his result but find that $f(q^2)$
is free from the zero-mode contribution if the denominator $D$ in
Eq.~(\ref{eq1}) contains a term proportional to the LF energy
$(k^-)^{n}$ with a power $n>0$.  The phenomenologically accessible
light-front quark model satisfies this condition $n>0$.

As we have shown in detail in Ref.~\cite{CJnew}, we can determine the
existence/nonexistence of the zero-mode contribution to $f(q^2)$ by
counting the powers of the LF energy $k^-$ in the denominator. For
example, if $D=D_{\rm cov}(k\cdot P)\equiv [2 k\cdot P + M_V(m_q+
m_{\bar q}) -i\epsilon]/M_V$, where $P$ is the four momentum of the
vector meson, see Ref.~\cite{MF97}, then $D$ contains a term proportional to the
LF energy $(k^-)^{n}$ with the power $n = 1$.  This power-counting
shows that the form factor $f(q^2)$ should not receive a zero-mode
contribution in the $D_{\rm cov}(k\cdot P)$ case.  We have confirmed
that the results found our way coincide with the ones from the
manifestly covariant calculation, while Jaus's method of finding
zero-mode contributions has a limitation in the choice of
$\Gamma^\mu$.

\end{document}